\documentstyle[prl,aps,preprint]{revtex}
\begin{document}
\title{Universalities of Triplet Pairing in Neutron Matter}
\author{V. A. Khodel, V. V. Khodel, and J. W. Clark}
\address{McDonnell Center for the Space Sciences
and Department of Physics,\\
Washington University, St. Louis, MO 63130 USA}
\date{\today , Submitted to Physical Review Letters}\maketitle
\begin{abstract}
 
The fundamental structure of the full set of solutions of the 
BCS $^3{\rm P}_2$ pairing problem in neutron matter is established.
The relations between different spin-angle components in these 
solutions are shown to be practically independent of density,
temperature, and the specific form of the pairing interaction.
The spectrum of pairing energies is found to be highly degenerate.
\end{abstract}

\vskip 1cm

Since the discovery of superfluidity in liquid $^3$He \cite{osh}, 
great advances have been made in understanding the properties of 
superfluid systems with triplet pairing.  In addition to the well-studied
case of liquid $^3$He below 2.6 mK~\cite{and,mermin,has,wol,vol}, 
triplet pairing is expected to occur in neutron matter in the quantum 
fluid interior of a neutron star~\cite{pin,tak,ost,bal,oslo}.  Neutrino 
cooling processes are strongly affected by $^3$P$_2$ pairing in this 
region \cite{page}, as is the vortex structure of the star and 
the coupling between core and crust \cite{sauls,lamb}.  A number 
of common features of superfluid, thermodynamic, and magnetic 
properties of different pair-condensed systems have been revealed by 
analyses based on symmetry principles \cite{has,wol,vol}, and further 
analytical insights have been gained near the critical temperature $T_c$
by application of the Ginzburg-Landau approach \cite{mermin}.  However, 
new universalities of triplet pairing may be uncovered by a direct 
attack on the BCS gap equation, as we shall now demonstrate.

The purpose of this letter is to identify fundamental solutions of the
triplet pairing problem in neutron matter and elucidate their structure
and their relationships.  If two identical spin-${1\over 2}$ fermions 
are paired with a nonzero total momentum ${\bf J}={\bf L}+{\bf S}$, 
the ordinary $S$-wave gap equation is converted into a system 
of coupled integral equations.  In the standard notation \cite{tak,ost}, 
we have the expansion 
$\Delta({\bf p}) =\sum\Delta_{LJ}^M(p)G_{LJ}^M({\bf n})$ 
of the gap in the spin-angle matrices 
$G_{LJ}^M({\bf n};s_1,s_2)=\sum C^{1M_S}_{{1\over 2}
{1\over 2}s_1s_2}C^{JM}_{1LM_sM_L}Y_{LM_L}({\bf n})$, 
with $\left(\Delta_{LJ}^M(p)\right)^*=(-1)^{J-M}\Delta_{LJ}^{-M}(p)$ assuming
time-reversal invariance.  The particle-particle interaction has the 
corresponding expansion $V({\bf p},{\bf p}')=\sum\langle p|V_{LJ}^{L'J}|p
\rangle G_{LJ}^M({\bf n})G_{L'J}^{M*}({\bf n}')$, with $|L-L'|\leq 2$ 
in the case of tensor forces.  The generalized BCS system then 
reads \cite{ost}
\begin{equation}
\Delta_{LJ}^M(p)= \sum_{L'L_1J_1M_1}(-1)^{\Lambda} \int 
  \langle p|V_{LJ}^{L'J}|p'\rangle S^{MM_1}_{L'JL_1J_1}({\bf n}')
  {\Delta_{L_1J_1}^{M_1}(p')\over 2E({\bf p}')}  \tanh {E({\bf p}') \over 2T}
  {\rm d}\tau' \, ,
\end{equation}
where $\Lambda=L-L_1+1$, $ {\rm d} \tau= p^2{\rm d}p{\rm d}{\bf n} 
\equiv {\rm d} \tau_0 {\rm d}{\bf n}$, 
and $S^{MM_1}_{LJL_1J_1}({\bf n})={\rm Tr}
\left[G_{LJ}^{M*}({\bf n})G_{L_1J_1}^{M_1}({\bf n})\right]$ 
accounts for the summation over spin variables.  The energy denominator
$E({\bf p})=[\xi^2(p)+D^2({\bf p})]^{1/2}$ involves the single-particle 
excitation energy $\xi(p)$ of the normal system and a gap function
whose square is constructed as
\begin{equation}
 D^2({\bf p})={1\over 2}\sum_{LJML_1J_1M_1}\Delta_{LJ}^{M*}(p)
 \Delta_{L_1J_1}^{M_1}(p) S^{MM_1}_{LJL_1J_1}({\bf n}) \, .
\end{equation}
The pairing parameter $\Delta$ measuring the gap value at the Fermi 
surface is given by 
$\Delta^2 = \int D^2(p_F{\bf n}) {\rm d} {\bf n}/{4\pi}$.

Due to the nonlinearity of the gap equation (1), one must in general
deal with off-diagonal effects in both the total ($J,J_1$) and 
the orbital ($L,L',L_1$) angular momentum quantum numbers.
However, for the present we follow the usual practice dating back 
to Ref. \onlinecite{and} and suppress these effects, thus allowing 
for superposition of spin-angle components only in the magnetic 
quantum number $M$. The analysis is greatly facilitated by a 
generalization of the separation method developed for S-wave 
pairing in Ref.~\onlinecite{kkc}.  Thus, defining 
$\phi_{LJ}(p)=\langle p|V_{LJ}^{LJ}|p_F\rangle /v_F$ and 
$v_F=\langle p_F|V_{LJ}^{LJ}|p_F\rangle $, we employ the decomposition
\begin{equation}
  \langle p| V_{LJ}^{LJ}| p'\rangle 
= v_F\phi_{LJ}(p)\phi_{LJ}(p')+W_{LJ}(p,p') 
\end{equation}
of the relevant pairing matrix into a separable portion and a
remainder $W_{LJ}(p,p')$ that vanishes identically when either
argument is on the Fermi surface.  Integrals containing $W_{LJ}$ 
as a factor are guaranteed to receive their overwhelming 
contributions some distance from the Fermi surface.
In such integrals, the replacements $E(p,{\bf n}) \rightarrow 
|\xi(p)|$ and $\tanh (E/2T) \rightarrow 1$ are justified to
high accuracy, errors in neutron matter being of relative order
$D^2({\bf p})/\epsilon_F^2 \sim 10^{-6}$, where $\epsilon_F$ is the 
Fermi energy.

Substituting (3) into (1) and invoking the orthogonality relation
$\int S_{LJLJ}^{MM_1}({\bf n}) {\rm d} {\bf n} = \delta_{MM_1}$,
the gap equations are recast as
\begin{equation}
  \Delta_{LJ}^M(p)+\int {W_{LJ}(p,p')\over 2|\xi(p')|}
  \Delta_{LJ}^M(p'){\rm d}\tau_0'=v_FB_{LJ}^M\phi_{LJ}(p) \, ,
\end{equation}
\begin{equation}
  B^M_{LJ}
= -\sum_{M_1} \int \phi_{LJ}(p)S^{MM_1}_{LJLJ}({\bf n})
  {\Delta_{LJ}^{M_1}(p)\over 2E({\bf p})}
  \tanh {E({\bf p})\over 2T} {\rm d} \tau \, .
\end{equation}
The quantities $B^M_{LJ}$ are merely numerical factors.  Consequently,
the $p$ dependence of all gap components is seen to be identical.
Specifically, we may write $\Delta_{LJ}^M(p)=D_{LJ}^M\chi_{LJ}(p)$, 
where the shape factor $\chi_{LJ}(p)$ obeys an integral equation 
\begin{equation}
   \chi_{LJ}(p) + \int W_{LJ}(p,p')
   {\chi_{LJ}(p') \over 2|\xi(p')|} {\rm d} \tau_0'=\phi_{LJ}(p) 
\end{equation}
of the same form as in the singlet case \cite{kkc}.  To determine
the amplitude $D_{LJ}^M$, we note that $\chi_{LJ}(p_F)=\phi_{LJ}(p_F)=1$  
since $W_{LJ}(p_F,p')=0$.  Therefore $\Delta_{LJ}^M(p_F)=D_{LJ}^M$,
and Eq.~(4) implies $D_{LJ}^M=v_FB^M_{LJ}$ while Eq.~(5) gives
\begin{equation}
   D^M_{LJ}
= -v_F\sum_{M_1} \int \phi_{LJ}(p)S^{MM_1}_{LJLJ}({\bf n})
  D^{M_1}_{LJ} {\chi_{LJ}(p)\over 2E({\bf p})} 
  \tanh {E({\bf p})\over 2T} {\rm d} \tau \ .
\end{equation}
The system (6)--(7) is more convenient for solution than the 
original equations (1), since the problem has been divided into
(i)~evaluation of the $M$-independent shape factor 
$\chi_{LJ}(p)$ from the nonsingular linear integral equation (6), and
(ii)~determination of the structure coefficients $D^M_{LJ}$ from the
nonlinear equation (7), where the log-singularity has been isolated.

Henceforth we specialize to the case $L=S=1$, $J=2$, this being 
the most favored uncoupled channel for pairing in neutron matter 
at densities prevailing in the quantum fluid interior of a neutron
star ($k_F=p_F/\hbar\sim 2$~fm$^{-1}$), where the $^1$S$_0$ gap has already 
closed \cite{tak,kkc}.  The arguments are simplified if we adopt 
$D^0_{12}\equiv\delta$ as a scale factor, write 
$D^{M\neq 0}_{12}\equiv (\lambda_M+i\kappa_M)\delta/\sqrt{6}$, and 
introduce a ``structure function'' 
\begin{eqnarray}
    d^2({\bf n}) = 
&& 16\pi D^2({\bf p})/\chi_{12}^2(p)
   = \delta^2[(1+\lambda_2)^2+\kappa^2_1 +\kappa^2_2
     +(\lambda_1^2-4\lambda_2-\kappa^2_1)x^2 \nonumber\\
&-& 2(\lambda_1+ \lambda_1\lambda_2+\kappa_1\kappa_2) xz 
    +(3+\lambda^2_1-\lambda^2_2-2\lambda_2)z^2\nonumber\\
&+& 2(2\kappa_2-\kappa_1\lambda_1)xy
    +2(\kappa_1+ \lambda_1\kappa_2-\lambda_2\kappa_1)yz] \, ,
\end{eqnarray}
where $x=\sin\theta\cos\varphi$, 
$y=\sin\theta\sin\varphi$, and $z=\cos\theta$.  After separation 
of the real and imaginary parts of the $D^{M\neq 0}_{12}$ in Eq.~(7), we 
arrive at the set of equations
\begin{eqnarray}
    \lambda_2
&=& -v_F[\lambda_2(J_0+J_5)-\lambda_1 J_1 -\kappa_1J_2-J_3]\, ,\nonumber \\
    \kappa_2
&=& -v_F[\kappa_2(J_0+J_5)-\kappa_1 J_1 +\lambda_1J_2+J_4] \, ,\nonumber\\
    \lambda_1
&=& -v_F[\lambda_1J_6-(\lambda_2+1)J_1+\kappa_2J_2-\kappa_1J_4/2]\, ,\nonumber\\
     \kappa_1
&=& -v_F[\kappa_1J_7-\kappa_2J_1-(\lambda_2-1)J_2-\lambda_1J_4/2]\, ,\nonumber\\
1 &=&-v_F[-(\lambda_1 J_1-\kappa_1 J_2+\lambda_2 J_3-\kappa_2 J_4)/3+J_5] \,,
\end{eqnarray}
which, in angular content, is consistent with the corresponding set 
in Ref.~\onlinecite{ost}.  The integrals $J_k$ are given by 
$J_6=(J_0+4J_5+2J_3)/4$, $J_7=(J_0+4J_5-2J_3)/4$, and, for 
$k=1,\cdots,5$, by
\begin{equation}
  J_k = \int f_k(\theta,\varphi){\phi_{12}(p)\chi_{12}(p) \over 
  2E({\bf p})}\tanh{E({\bf p})\over 2T} {\rm d} \tau \, ,
\end{equation}
with
$f_0=1-3z^2$, $f_1=3xz/2$, $f_2=3yz/2$, $f_3=3(2x^2+z^2-1)/2$, $f_4=3xy$, 
and $f_5=(1+3z^2)/2$.

The system (9) has three one-component solutions \cite{tak,ost} with 
$|M|=0,~1$, and 2.  We preface our analytic exploration of multicomponent 
solutions with the following observation.  Substitution of 
$\partial d^2(\theta,\varphi)/\partial\varphi$ for $f_k$ in 
definition (10) must yield zero upon integration over $\varphi$.  This 
identity implies a relation 
\begin{equation}
  \sum_{k=1}^4 c_kJ_k=0
\end{equation}
between the $J_k$ integrals,
with $c_1=\kappa_1+\lambda_1\kappa_2- \lambda_2\kappa_1$, 
$c_2=\lambda_1+\lambda_1\lambda_2+\kappa_1\kappa_2$, 
$c_3=2\kappa_2-\kappa_1\lambda_1$, and 
$c_4=2\lambda_2-(\lambda_1^2-\kappa_1^2)/2$.
Now observe that if the first equation of (9) is multiplied
by $2\kappa_2$, the second by $2\lambda_2$, the third by $\kappa_1$, 
and the fourth by $-\lambda_1$, and the results of the last
three operations are subtracted from that of the first, relation 
(11) is reproduced.  Thus only four of the five equations in (9) 
are truly independent and hence any one of the parameters 
$\lambda_1$, $\lambda_2$, $\kappa_1$, $\kappa_2$ 
can be chosen arbitrarily.  We take $\kappa_1=0$.  
With this choice, solutions of (9) are necessarily even functions of 
$\lambda_1$, so attention may be focused on the sector $\lambda_1\geq 0$. 

The search for multicomponent solutions begins with the
restricted case $\kappa_2=0$, for which $d^2({\bf n})$ is independent 
of $y$. In Eq.~(10), the integration of $f_k(\theta,\varphi)$ over $y$ 
is then carried out using the formula 
$\sin\theta d\theta d\varphi =2\delta(x^2+y^2+z^2-1)dx~dy~dz$.  
For $k=1,3$, this yields $2f_k(x,z)(1-x^2-z^2)^{-1/2}$ 
since $f_1$ and $f_3$ are independent of $y$, while 
both $J_2$ and $J_4=0$ vanish because $f_2$ and $f_4$ are odd in $y$ 
and the $y$ integral has symmetric limits.  As a result, 
there remain only three independent equations,
\begin{eqnarray}
\lambda_2 &=& -v_F[\lambda_2(J_0+J_5) -\lambda_1 J_1-J_3] \, ,\\
\lambda_1 &=& -v_F[\lambda_1(J_0/4+J_5)-(\lambda_2+1)J_1
              +\lambda_1 J_3/2] \, ,\\
        1 &=& -v_F[-(\lambda_1 J_1+\lambda_2 J_3)/3+J_5] \, .
\end{eqnarray}
We first identify and verify a particular solution with
$\lambda_1=0,\lambda_2=3$, for which the structure function (8) 
becomes $d^2(x,z)=4\delta^2[4-3(x^2+z^2)]$.  The symmetry of this 
function with respect to $x$ and $z$ implies the relation $3J_0+2J_3=0$, 
since the combination $3f_0+2f_3=6(x^2-z^2)$ changes sign on interchange 
of $x$ and $z$ whereas $d^2$ and other factors within the integrand of 
(10) are left unchanged.  Further, $J_1(\lambda_1=0,\lambda_2=3)=0$
since $f_1=xz$ is an odd function of $x$.  Under these conditions,
Eq.~(13) is satisfied identically, while Eqs.~(12) and (14) coincide and 
the resulting equation, $1=-v_F(J_5-J_3)$, determines $\delta$.  All other 
solutions of the set (9) are more degenerate.  To illustrate this important 
feature, let us put $\lambda_2=-1$.  Then, at {\it any} $\lambda_1$ the 
structure function (8) is seen to take the factorized form 
$d^2({\bf n};\lambda_1,\lambda_2=-1)=\delta^2(\lambda_1^2+4) (x^2+z^2)
= 24\pi\Delta^2(x^2+z^2)$.  The symmetry of $d^2$ in $x$ and $z$ again 
implies the relation $3J_0+2J_3=0$, while integration of $f_1=xz$ over 
$x$ gives 0 and therefore $J_1=0$.  It follows that Eqs.~(12)--(14) 
again coincide but now provide an equation $1=-v_F[J_3/3+J_5]$ that 
determines $\Delta^2$ rather than $\lambda_1$ or $\delta$ individually.  
Here we have a striking example of the universal structure of solutions 
of the $^3{\rm P}_2$ pairing problem, also manifested in the remaining 
solutions of the system (9).

These further solutions are found by implementing a rotation 
$R=\bigl( x=t\cos\alpha+u\sin\alpha,\, z=t\sin\alpha-u\cos\alpha\bigr)$. 
Expressing $d^2$ in terms of $t$ and $u$ and setting $\tan \alpha = \gamma$, 
one easily finds conditions
\begin{equation}
  (\lambda^2_1-4\lambda_2)\gamma^2
+ 2\lambda_1(1+\lambda_2)\gamma
+ \lambda^2_1-\lambda^2_2-2\lambda_2 +3=0 \, ,~~~
  \lambda_1\gamma^2-(\lambda_2-3)\gamma -\lambda_1=0 
\end{equation}
under which $d^2$ becomes a function of $t$ only.  The choice 
$\gamma(\lambda_1,\lambda_2)=\gamma_0(\lambda_1,\lambda_2) =\\
\lambda_1(1+\lambda_2)/(4\lambda_2-\lambda^2_1)$ 
meets both conditions provided
\begin{equation}
(\lambda^2_1-2\lambda_2+2)(\lambda^2_1-2\lambda^2_2-6\lambda_2)=0 \, .
\end{equation}
Equation~(16) embodies three branches of $\lambda_2$ versus $\lambda_1$, 
which start as parabolas from $\lambda_1=0$ and $\lambda_2=1$, 0, and $-3$.  
The structure function has $t$ dependence $d^2(t)\propto{1-t^2}$ 
when the first factor of (16) vanishes and 
$d^2(t)\propto{1+3t^2}$ when the second is zero.
Calculating the integrals $J_1$ and $J_3$ by rotation of the 
$x,z$ plane under $R$, we are led to the relations
\begin{eqnarray}
    (\lambda^2_1-4\lambda_2)J_1+\lambda_1(\lambda_2+1)J_3
&=& 0 \, ,\nonumber\\
    3\lambda_1(1+\lambda_2)J_0-2(\lambda^2_1-2\lambda^2_2+6)J_1
&=& 0 \, .
\end{eqnarray}
The first relation (for example) is verified as follows, noting
that the integrand on its l.h.s. is proportional to 
$[(\lambda^2_1-4\lambda_2)f_1+\lambda_1(\lambda_2+1)f_3]/(1-t^2-u^2)^{1/2}$. 
Substituting $f_1(t,u)$ and $f_3(t,u)$ and integrating over $u$, 
which can be done freely for any shape of $d^2(t)$, 
we obtain a result that is proportional to 
$(\lambda^2_1-4\lambda_2)\gamma+\lambda_1(\lambda_2+1)$ and therefore
vanishes when $\gamma_0(\lambda_1,\lambda_2)$ is substituted.  What is 
remarkable is that Eqs.~(12)--(14) coincide when relations (17) are 
inserted, and once again these equations determine only $\Delta^2$.
Thus, construction of the rotation $R$ ``kills two birds with one stone'': 
the condition (16) required to transform $d^2$ into one-dimensional 
form also specifies another set of solutions of our system.  Within 
the constraint $\kappa_1=0$, these solutions possess a line degeneracy 
as opposed to the point (or nondegenerate) character of the solution 
$(\lambda_1 = 0, \lambda_2 = 3)$.  In addition to the free choice made 
for $\kappa_1$, one of the coefficients $\lambda_i$ can be assigned 
arbitrarily.

We now allow $\kappa_2$ to have a nonzero value, thus bringing in
the second of Eqs.~(9).  At $\lambda_1=0$, this equation becomes identical
with the first of the set, as is seen with the aid of Eq.~(11).  
The particular solution $(\lambda_1=0,~\lambda_2=3)$ is then replaced 
by one with $\lambda_1=0$ and $(\lambda_2^2+\kappa_2^2)^{1/2}=3$, but 
the pairing energy remains unaltered, the relevant quantities being
independent of the phase of the coefficient $D^2_{12}(\lambda_1=0)$. 
To find the other multicomponent solutions in the general case with 
$\kappa_2\neq 0$ and $\lambda_1\neq 0$, we may extend our previous 
tactic and apply a rotation in {\it three}-dimensional
space so as to eliminate four terms in expression (8) and cast
$d^2$ into a one-dimensional form.  The three Euler angles are 
thereby fixed, implying the single relation 
\begin{equation}
  \kappa^2_2=(1+\lambda_2)(\lambda^2_1/2-\lambda_2+1) 
\end{equation}
between $\lambda_2$,
$\kappa_2$, and $\lambda_1$, 
which can be shown to satisfy all of Eqs.~(9).  This relation defines
two branches  $\kappa_2(\lambda_1,\lambda_2)$.  Starting at the plane 
$\kappa_2=0$, one branch grows out of the solution $\lambda_2=-1$ 
while the other grows out of the parabola $\lambda^2_1/2=\lambda_2-1$ 
contained in Eq.~(16).  (Accordingly, $\lambda_2=-1$ and this 
parabola cannot be counted as independent solutions.)  The two 
surfaces defined by (18) complete the set of states of the 
$^3{\rm P}_2$ problem.

The solutions we have identified divide into two groups, the states 
within a group being essentially degenerate in energy.  This 
behavior is consistent with the numerical calculations reviewed in 
Ref.~\onlinecite{tak}.  The group with lowest energy, having 
structure function $d^2(t)\propto{1+3t^2}$, contains only 
nodeless states and consists of (i)~the particular state 
($\lambda_1=0$, $\sqrt{\lambda_2^2+\kappa^2_2}=3$) and (ii)~the states 
belonging to the branches of Eq.~(16) starting at the points 
$(\lambda_1,\lambda_2)=(0,0)$ and $(0,-3)$.  The upper group contains 
the remaining states, having $d^2(t)\propto{1-t^2}$ and one node.
The splitting between the two groups can be calculated (for example) 
as the splitting between the states $\lambda_2=-1$ and 
($\lambda_1=0$, $\lambda_2=3$), henceforth labeled ${\rm u}$ and 
${\rm l}$ respectively.  Evaluating (14) for both states, one 
finds the relation $J_3^{({\rm u})}/3+
J_3^{({\rm l})}+J_5^{({\rm u})} -J_5^{({\rm l})} =0$ 
between integrals of the form (10).  In explicating 
this relation, we exploit the fact that the dominant contributions 
to the integrals $J_3^{({\rm u})}$, $J_3^{({\rm l})}$, and
$J_5^{({\rm u})}-J_5^{(\rm u}$ come from the range of $p$ values 
adjacent to the Fermi surface, where 
$\chi_{12}(p)$ and $\phi_{12}(p)$ are effectively unity.
One readily arrives at the analytical result
\begin{equation}
 \ln{{\Delta_{\rm u}^2} \over {\Delta_{\rm l}^2}} (T=0)
  ={2\pi\over 9\sqrt{3}}+{2\over 3}-\ln 3\simeq -0.028 
\end{equation}
for the splitting of upper and lower states, in close agreement with 
Ref.~\onlinecite{tak}.  Similar results are also available at 
finite temperature $T$.  

The conclusions that follow from these exercises are that if the 
mixing of different $L,J$ channels is neglected, (i) the $^3{\rm P}_2$ 
gap spectrum is nearly degenerate and (ii) its structure, in terms of 
energy splittings between the different states, is a universal function 
of $T/T_c$, independent of any other input parameters including 
the density.  In particular, the concrete form of the particle-particle 
interaction $V$ was not used anywhere, so the structure and relations 
we have established retain their validity even when fluctuation and 
polarization corrections to the bare $V$ are taken into account.

Finally, we return to the issue of nondiagonal contributions to
the system (1) of gap equations, which arise principally from the
$^3$P$_2$--$^3$F$_2$ coupling, and outline a perturbative evaluation 
\cite{and} of their effects.  The r.h.s. of each equation of the set 
(9) is now perturbed by a small ``nondiagonal'' contribution.  In the 
presence of these additional terms, the degeneracies found above 
are removed.  The pairing energy no longer depends on $d^2({\bf n})$ alone, 
and the parameters $\lambda_i,\kappa_i$ are fully determined.  Consider, 
for example, the alteration of the parabolic branch contained in the 
relation (16), which may be measured by a new variable 
$\zeta=\lambda^2_1-2\lambda_2+2$.  
Both $\zeta$ and the change $\eta=\delta(\zeta)-\delta(\zeta=0)$ of the 
scale factor $\delta$ are expected to be small; therefore in performing  
Taylor expansions of the $J_k(\zeta,\eta,\lambda_1)$ we need only 
retain terms linear in $\zeta$ or $\eta$.  The original set of three
equations (12)--(14) is replaced by three new ones, each of which takes the
schematic linear form $\zeta A(\lambda_1)+\eta B(\lambda_1)=P(\lambda_1)$
with different choices of the functions $A$, $B$, and $P$, all referred
to $\zeta=\eta=0$.  The small quantities $\zeta$ and $\eta$ may be obtained 
from any pair of the equations, as functions of $\lambda_1$.  Substituting
these functions into the remaining equation, we arrive at a closed
form that determines the value of $\lambda_1$, which was hitherto
arbitrary.  Estimation and analysis of available numerical results 
\cite {tak,ost} indicate that the nondiagonal corrections to the 
universal relations we have derived for triplet pairing in neutron 
matter are small, maximally of order several percent of the
splitting given by Eq.~(19).

In summary, straightforward arguments based on a new separation 
method \cite{kkc} for treating BCS-type gap equations have revealed 
the structure and energetics of the full set of solutions of 
the pairing problem in the uncoupled $^3$P$_2$ channel. In contrast to 
the Ginzburg-Landau scheme employed by Mermin \cite{mermin}, the
present approach is applicable at any temperature $T$.  The analysis 
shows that the structure of the solutions is in fact universal --- 
independent of the temperature, the density, and the specific parameters 
of the interparticle potential, which affect only an overall 
scale factor in the pairing energies. The line of analysis we 
have followed transcends the problem considered here.  An obvious 
future objective is to characterize the solutions occurring 
in the problem of superfluid $^3$He.  The structure function 
$d^2(x,y,z)$ is again bilinear in its variables, but the states 
for $L=S=1$ with $J=0,1,2$ contribute on an equal footing and the 
number of equations in the system analogous to (9) rises from five 
to nine.  The treatment introduced herein could also be relevant to the 
description of superdeformed bands in atomic nuclei, if
triplet P-wave pairing is responsible for this phenomenon as 
suggested in Ref.~\onlinecite{shap}.

We acknowledge helpful discussions with R.~Fisch, P.~Schuck, and 
G.~E.~Volovik.  This research was supported by the Russian 
Foundation for Basic Research under Grant No.~96-02-19292 and by 
the U. S. National Science Foundation under Grant No.~PHY-9602127.


\begin{thebibliography} {99}
\bibitem{osh} 
D.~D.~Osheroff, R.~C.~Richardson, and D.~M.~Lee, 
Phys. Rev. Lett. {\bf 28}, 885 (1972). 
\bibitem{and}
P.~W.~Anderson and P.~Morel, Phys. Rev. {\bf 123}, 1911 (1961). 
\bibitem{mermin}
N. D. Mermin, Phys. Rev. {\bf A9}, 868 (1974). 
\bibitem{has} 
Y.~Hasegawa, T.~Usagawa, and F.~Iwamoto, 
Prog. Theor. Phys. {\bf 62}, 1458  (1979).
\bibitem{wol} 
D.~Vollhardt and P.~W\"olfle, {\it The Superfluid Phases of Helium 3} 
(Taylor \& Francis, London, 1990).
\bibitem{vol} 
G.~E.~Volovik, in {\it Helium Three}, ed. W.~P.~Halperin
and L.~P.~Pitaevskii (North Holland, Amsterdam, 1990).
\bibitem{pin} 
D.~Pines and M.~A.~Alpar, Nature {\bf 316}, 27 (1985).
\bibitem{tak} 
T.~Takatsuka and R.~Tamagaki, 
Prog. Theor. Phys. Suppl. {\bf 112}, 27 (1993). 
\bibitem{ost} 
L.~Amundsen and E.~\O stgaard, Nucl. Phys. {\bf A442}, 163 (1985). 
\bibitem{bal} 
M.~Baldo, J.~Gugnon, A.~Lejeune, and U.~Lombardo, 
Nucl. Phys. {\bf A515}, 409 (1990).
\bibitem{oslo} 
{\O}. Elgar{\o}y, L. Engvik, M. Hjorth-Jensen, and E. Osnes,
Nucl. Phys. {\bf A607}, 425 (1996).
\bibitem{page}
D. Page and J. H. Applegate, Ap. J. Lett. {\bf 394}, L17 (1992).
\bibitem{sauls}
J. A. Sauls, D. L. Stein, and J. W. Serene, Phys. Rev. D {\bf 25}, 967
(1982).
\bibitem{lamb}
F. K. Lamb, in {\it Frontiers of Stellar Evolution}, 
ed. D. L. Lambert (Astronomy Society of the Pacific, 
San Francisco, 1991), p. 299.
\bibitem{kkc} 
V.~A.~Khodel, V.~V.~Khodel, and J.~W.~Clark, 
Nucl. Phys. {\bf A598}, 390 (1996).
\bibitem{shap} V.~I.~Fal'ko and I.~ S.~Shapiro, 
Sov.\ Phys.\ JETP {\bf 64}, 706 (1986).

\end{thebibliography}
\end{document}